\documentclass{PoS}
\usepackage{amssymb}
\usepackage{fontenc}
\usepackage{times}
\usepackage{mathptmx}
\usepackage{graphicx}
\usepackage{subcaption}
\usepackage{amsmath}
\usepackage{caption}

\title{Production of pions, kaons and protons in Xe--Xe collisions at 
$\mathbf{\sqrt{{\textit s}_{\mathbf{\rm \mathbf{NN}}}}=5.44} \text{ TeV}$  }

\ShortTitle{Production of pions, kaons and protons in Xe--Xe at $\sqrt{s_{\rm 
NN}}=5.44\text{ TeV}$}

\author{\speaker{Simone Ragoni}, for the ALICE Collaboration\\
        Universit\`{a} di Bologna and INFN (Bologna)\\
        E-mail: \email{simone.ragoni@cern.ch}}

\abstract{In late 2017, the ALICE experiment recorded a  data sample of Xe--Xe 
collisions at the unprecedented energy in A--A systems of $\sqrt{s_{\rm{NN}}} = 
5.44\text{ TeV}$. The $p_{\rm T}$ spectra at mid-rapidity ($|y| < 0.5$) of 
pions, kaons and protons are presented. The preliminary $p_{\rm T}$ spectra are 
obtained by combining independent analyses with the Inner Tracking System 
(ITS), the Time Projection Chamber (TPC), and the Time-Of-Flight (TOF) 
detectors.
This paper focuses on the details of the analysis performed with TOF and in 
particular on the performance implications of the special Xe--Xe run 
conditions. The peculiarity of this data set comes from the experimental 
conditions: because of the lower magnetic field of
the ALICE solenoid ($B = 0.2\text{ T}$, lower than 
the nominal  $0.5\text{ T}$) we expect to explore a $p_{\rm T}$ region  
unattainable before. A comparison between the yields at different centrality 
bins will also be provided. }

\FullConference{Sixth Annual Conference on Large Hadron Collider Physics (LHCP2018)\\
                4-9 June 2018\\
                Bologna, Italy}

\begin{document}

\section{Introduction}
The ALICE Collaboration has built a dedicated heavy-ion collisions detector 
to investigate a new state of 
matter, the QGP (Quark--Gluon Plasma), which is expected to be created in these 
collisions.
One way to quantify the expansion velocity of the system produced in heavy-ion 
collisions is to study the production of identified particles, particularly 
their transverse momentum distributions. The final-state distributions are 
governed by the temperature at which the system freezes out and the particles 
no longer interact. If the values of the temperature  agree for different 
particle species, e.g. pions, kaons, protons, and antiprotons, this indicates a 
global freeze-out temperature and fluid velocity. While the low-$p_{\rm T}$ end 
of the spectrum involves particles which are more likely to be in thermal 
equilibrium, the high-$p_{\rm T}$ particles are 
more likely to have been produced in hard scatterings, governed by perturbative 
QCD.

Ultimately, this study addresses the question whether the matter reaches kinetic 
equilibrium. If we suppose that the system created in heavy-ion collisions is in kinetic equilibrium, 
the pressure is 
built inside the system. The matter is surrounded by vacuum, so a pressure 
gradient in outward 
directions generates collective flow and, in turn, the system expands 
radially.  When the matter is 
moving at a finite velocity the momentum distribution is Lorentz boosted and 
the magnitude of this boost is proportional to the particle mass.  If this kind 
of effect in the
momentum distribution can be observed experimentally, one can obtain some information about kinetic 
equilibrium. Assuming each fluid element expands radially at a radial flow 
velocity $\beta_{\rm T}$ , the 
$p_{\rm T}$ spectra for pions and protons can be calculated by convoluting 
these affected momentum 
distributions over the azimuthal direction, which is realized in the 
\textit{blast wave model} 
\cite{checks_10, checks_11}. 

\section{Analysis procedure}
The measurement of the production of $\pi^{\pm}$, K$^{\pm}$, p and 
$\bar{\text{p}}$ in Xe--Xe collisions at $\sqrt{s_{\rm NN}}=5.44\text{ TeV}$ 
has been performed for ITS, TPC and TOF \cite{detector} separately, but in the 
following we 
will focus on the TOF analysis only.

The final goal is the combination of all individual analyses so as to provide a 
single spectrum covering the widest $p_{\rm T}$ interval possible with the best 
precision, combined results are shown in Sec.~\ref{spectra}. The final spectra 
represent the combination of the three individual analyses. The $p_{\rm 
T}$-ranges of 
the different analyses are given in Tab.~\ref{table:ptr}.

\begin{table}[h]
  \begin{center}
    \begin{tabular}{|c|c|c|c|}
      \hline
      \textbf{Analysis} & \bf $\pi$ & \bf K & \bf p\\ \hline
      \bf ITSsa & 0.08-0.70 GeV$/$\textit{c} & 0.20-0.45 GeV$/$\textit{c} & 0.30-0.50 GeV$/$\textit{c} \\ \hline
      \bf TOF & 0.40-5.00  GeV$/$\textit{c}  & 0.4-3.60 GeV$/$\textit{c} & 0.50-5.00 GeV$/$\textit{c} \\ \hline 
      \bf TPC & 0.25-0.70  GeV$/$\textit{c}  & 0.25-0.45 GeV$/$\textit{c} & 0.40-0.80 GeV$/$\textit{c} \\ \hline
    \end{tabular}
  \end{center}
  \caption{$\pi$, K and p $p_{\rm T}$ ranges in GeV$/$\textit{c} used for the 
  different detectors in
  the analysis of Xe--Xe collisions.}
  \label{table:ptr}
\end{table}
\subsection{TOF analysis}
The analysis relies on the global tracks and the extraction of the total yield 
by performing particle identification
(PID) exclusively with the TOF detector. The main goal of this analysis is to 
provide spectra as a function of $p_{\rm T}$ for identified
$\pi^+$, $\pi^-$, K$^+$, K$^-$, p and $\bar{\text{p}}$ at intermediate $p_{\rm 
T}$, 
which are then combined with the ITS and TPC spectra. 
The  strategy, as it will be better
explained in the following, is based on the comparison between the measured time of flight and the
\textit{theoretical prediction} for each mass hypothesis.

\begin{figure}
  \centering
  \begin{subfigure}{0.45\textwidth}
    \centering
    \includegraphics[width=\linewidth]{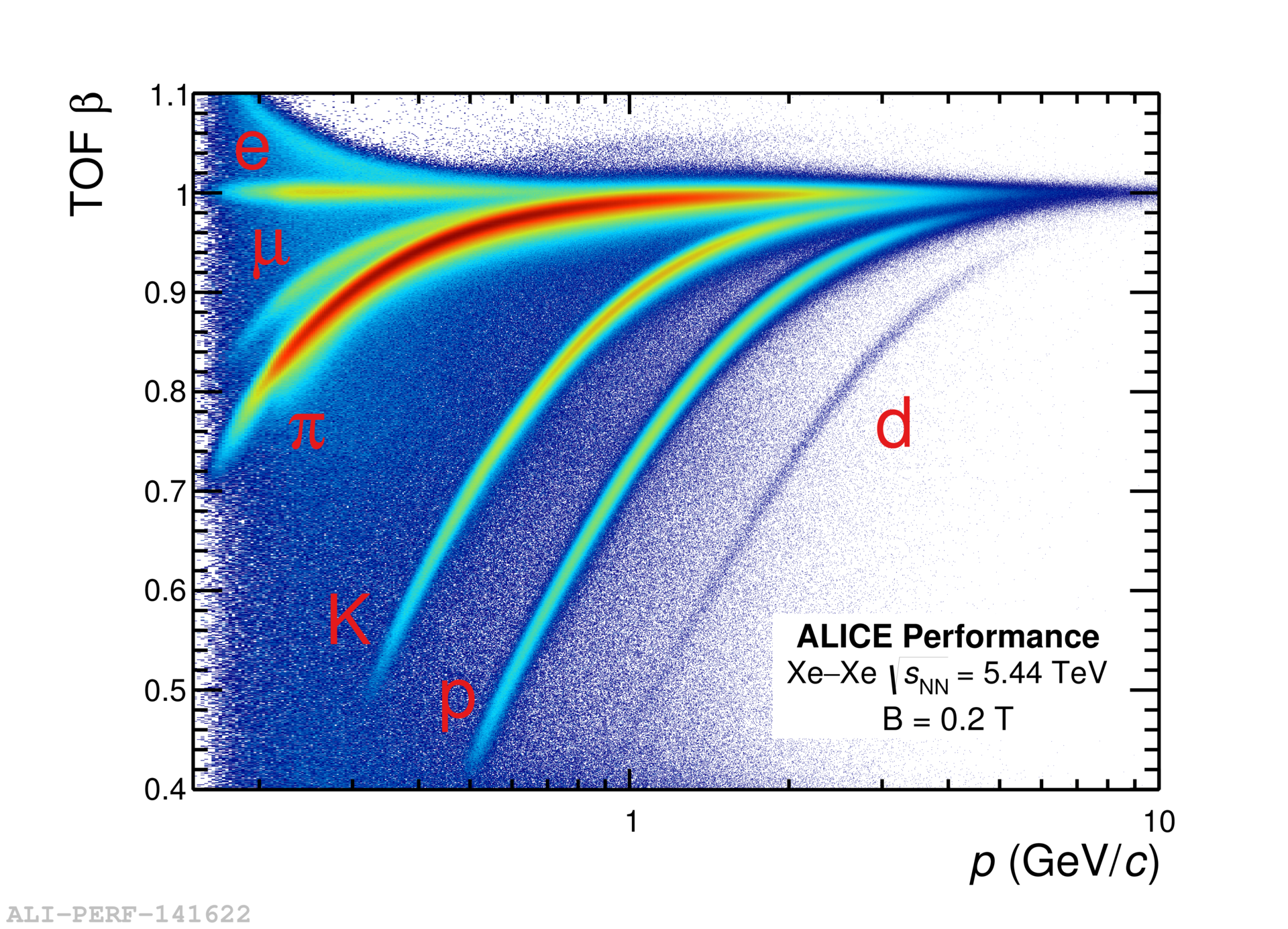}
    \caption{Beta vs $p$ measured with the TOF detector.}
    \label{fig:tof:TOFbetaxexe}
  \end{subfigure}
  \begin{subfigure}{0.45\textwidth}
    \includegraphics[width=\linewidth]{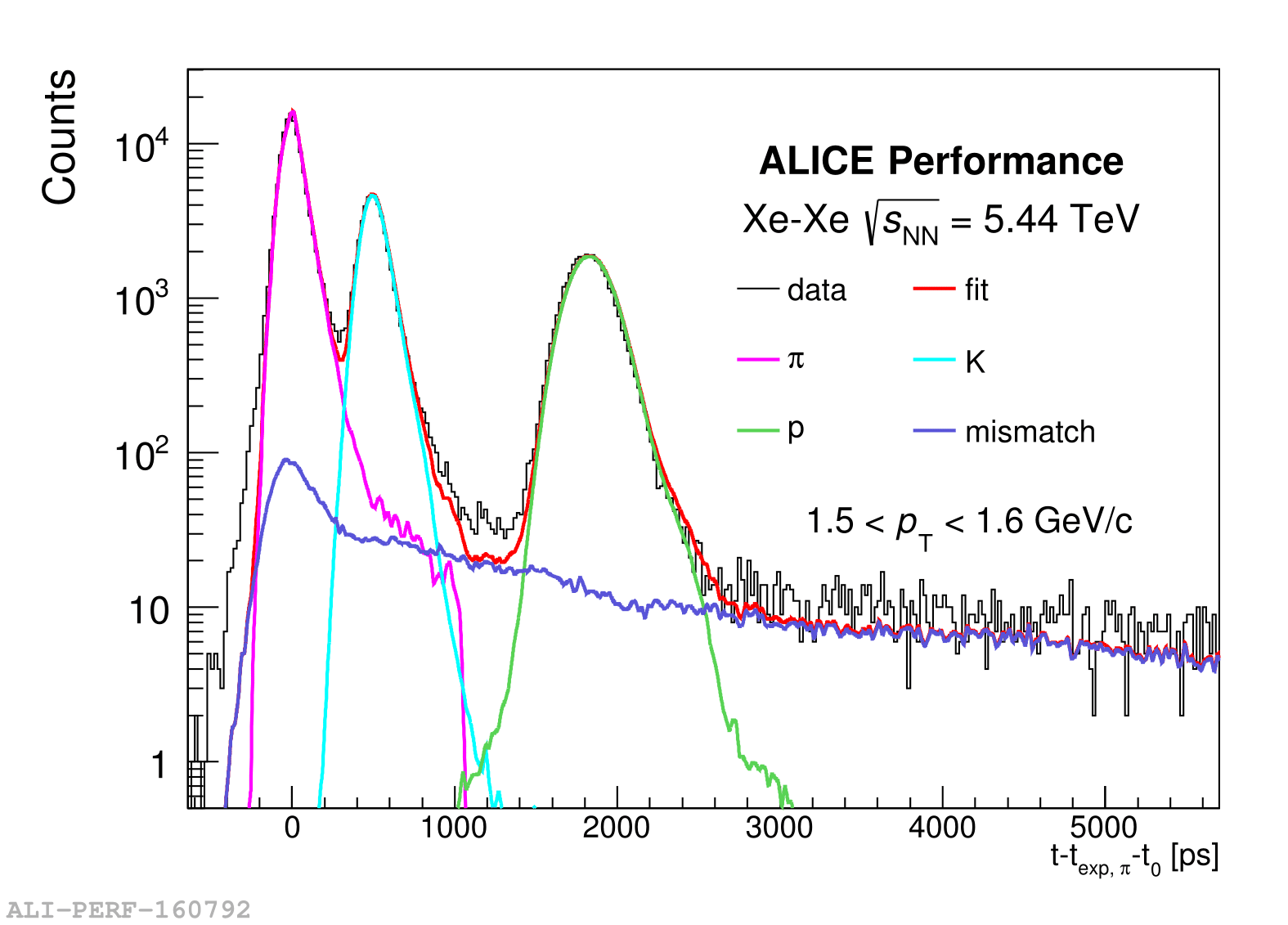}
    \caption{TOF signal: single template components.}
    \label{fig:TOF_templates}
  \end{subfigure}
    \caption{Left panel: beta measured with the TOF detector as a function of 
    the track momentum in Xe--Xe collisions at $\sqrt{s_ {\rm NN}} = 5.44\text{ 
    TeV}$. The visible bands are from $e$, $\mu$, $\pi$, K, p and d. The 
    contribution from wrongly associated tracks can be seen outside the bands. 
    Right panel: the TOF signal fit with single template components. The 
    mismatch is depicted with a dark blue color, pions with magenta, kaons 
    with light blue and protons with green.}
\end{figure}

The analysis strategy mainly relies on tracks reconstructed in the TPC, which 
are extrapolated to the TOF. 
The assignment of a TOF cluster to the propagated track allows one to use the precise measurement of the arrival time of the particle at the TOF surface.
Examples of the measured particle velocity can be seen in Fig. \ref{fig:tof:TOFbetaxexe} for Xe--Xe collisions.

The identification of different particle types is performed with a statistical 
approach, measuring the \textit{time-of-flight} as 
  \begin{equation} \label{eq:timeofflight}
    time\text{-}of\text{-}flight = T_{\rm TOF} - T_{\rm 0}\text{ .}
  \end{equation}
The time of flight is defined as the difference between the time measured by 
the TOF detector and the start time $T_{\rm 0}$.
For simplicity from now on we will refer to the $time\text{-}of\text{-}flight$ 
as $t$.
  
The  \textit{expected time-of-flight} ($t_{\rm exp, \text{\it i}}$) can be 
numerically 
computed for every particle species \textit{i} for each track by taking into 
account the 
track length and the energy loss in the material.

%
The PID strategy with TOF takes advantage of the separation between the 
different particles by calculating the $n\sigma$ separation
  \begin{equation} \label{eq:tofnsigma}
    n\sigma_{i} = \frac{t - t_{{\rm exp,}i}}{\sigma_{i}}\text{ ,}
  \end{equation}
indicating by $n \sigma_{i}$ the distance in number of sigmas of the measured 
value from the expected one under a specific mass hypothesis, and by 
$\sigma_{i}$ the uncertainty on the numerator. The uncertainty $\sigma_{i}$ can 
be expressed as 
$\sigma_{i}^2 = \sigma_{\rm TOF}^2 + \sigma_{T_{\rm 0}}^2 + \sigma_{t_{\rm 
exp}}^2$ which 
are the intrinsic contribution due to the TOF
itself, the resolution of the start time of the event, and the
contribution due to the expected time of flight for a certain
particle species, respectively.
  

The $T_0$ can be determined with different methods \cite{time_event}, each with 
different precision. The measurement of the $T_0$ with the TOF is fully 
efficient in Xe--Xe collisions from central collisions up to centrality bin 
$80$--$90\%$, where the multiplicity of tracks reaching the TOF is not 
high enough  to ensure the determination of the event time in each collision. 
In 
this case, the measurement of the start time relies mostly on the T0 detector 
or, in case it is not available, on the nominal bunch crossing time, which 
results in a worse \textit{time-of-flight} resolution.
\subsection{TOF signal}
\begin{figure}
  \centering
  \begin{subfigure}{0.66\textwidth}
      \centering
      \includegraphics[width=0.6\linewidth]{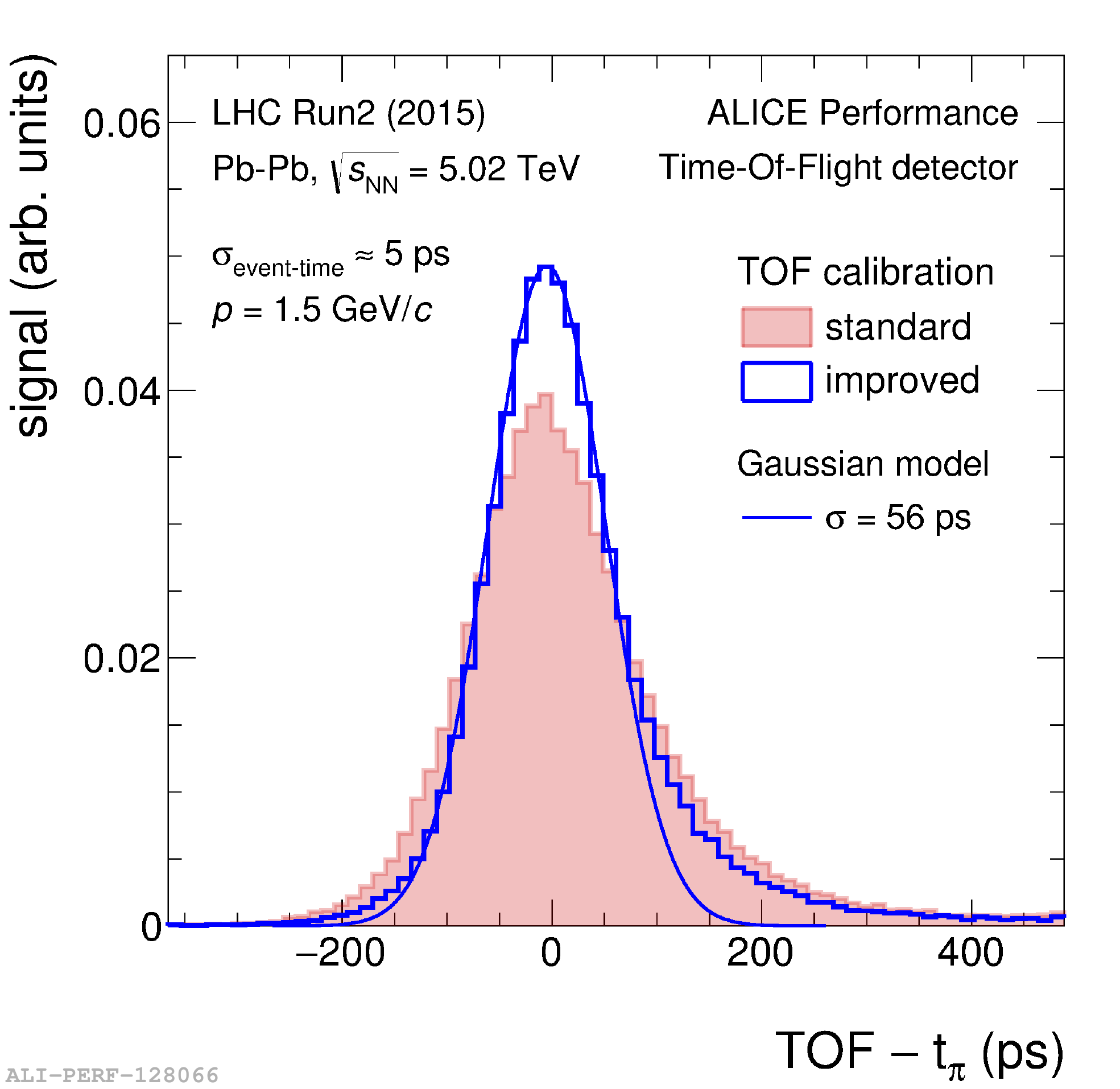}
      \caption{TOF signal.}
      \label{fig:tof_signal}
  \end{subfigure}
  \begin{subfigure}{0.66\textwidth}
    \centering
    \includegraphics[width=0.7\linewidth]{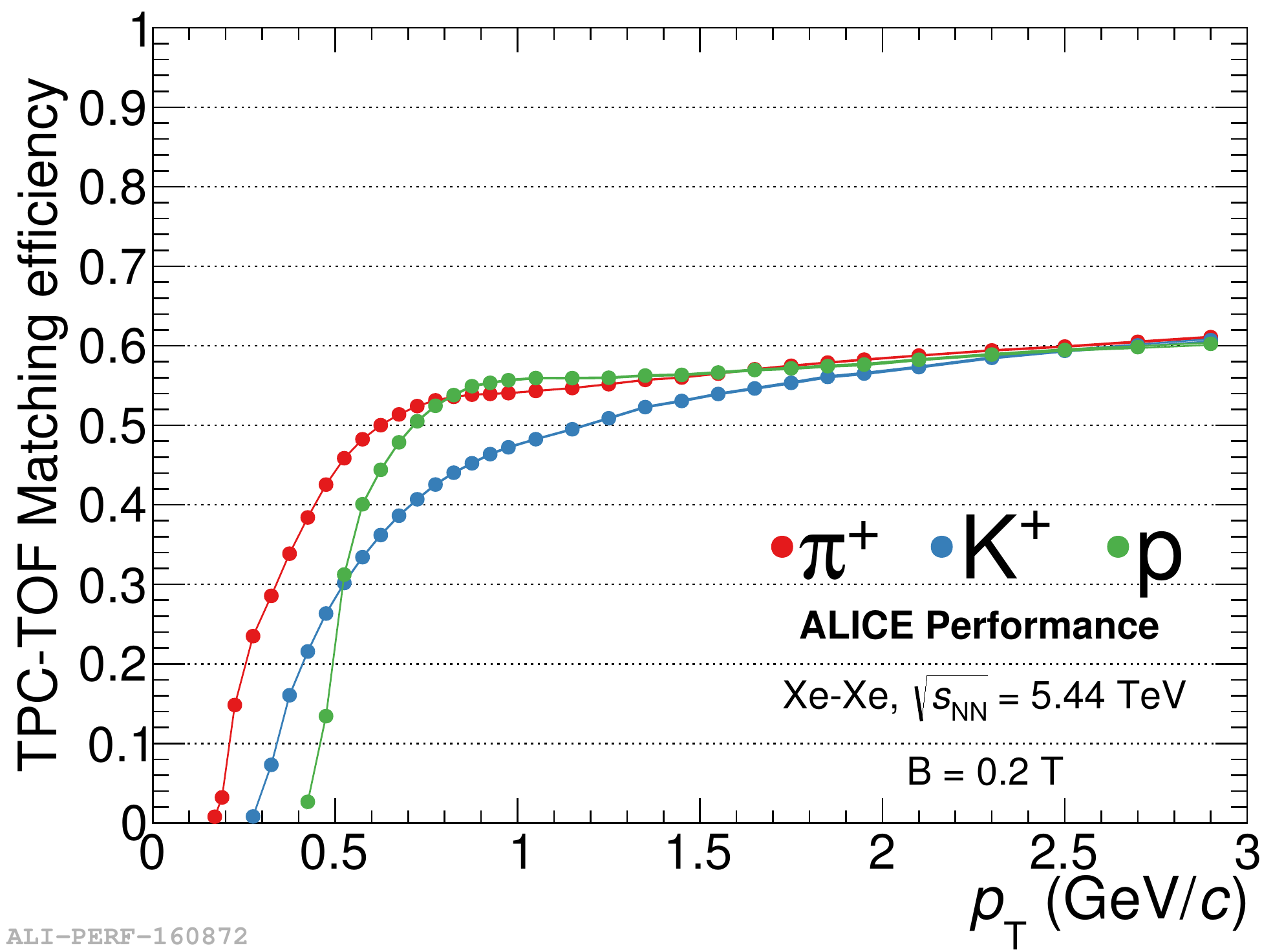}
    \caption{Matching efficiency.}
    \label{fig:match}
  \end{subfigure}
    \caption{The TOF signal (upper panel) and the matching efficiency for 
    pions, 
    kaons, and protons (lower panel).}
 \end{figure}

  The signal in the TOF detector must be correctly parametrized to perform the particle identification.
  The parameterization is given by
  \begin{equation} \label{eq:convolution}
    f(t)=\begin{cases}
      A\cdot \textit{Gaus}(t, \mu, \sigma), & \textit{if $t\leq \mu+\tau$}.\\
      A\cdot \textit{Gaus}(\mu+\tau, \mu, \sigma)\cdot 
      \exp\left[-\frac{\tau\cdot\left(t-\mu-\tau\right)}{\sigma^2}\right], & 
      \textit{otherwise}.
    \end{cases}
  \end{equation}
  
   where $A$ is a normalization factor.
  Basically the actual form of $f(t)$  is  a Gaussian distribution with an exponential tail on the right side of the peak, see Fig.~\ref{fig:tof_signal}.
  

  The parameters are tuned on experimental data and are set to $\mu = 0$ and 
  $\tau = 
  0.85 \sigma$.
  These parameters are intrinsic to the TOF detector and are to be taken as asymptotic values that are measured once the particle momentum is high enough that the energy loss becomes negligible.
  The parameters used to describe the TOF signal are extracted with a fit to 
  the distribution of  $t - t_{\rm exp,\pi} - t_{\rm 0}$ in the region where 
  the 
  peak for $\pi$ is clearly separated in TOF.

The raw spectra  extracted in this way have to be corrected for the presence 
of secondary 
particles, for the tracking efficiency, and the matching efficiency (see 
Fig.~\ref{fig:match}), before they can undergo combinations with spectra from 
ITS and TPC.
\section{Results}
\label{spectra}
\begin{figure}
  \centering
  \begin{subfigure}{0.3\textwidth}
      \centering
      \caption{Combined spectra for $\pi$.}
      \includegraphics[width=\linewidth]{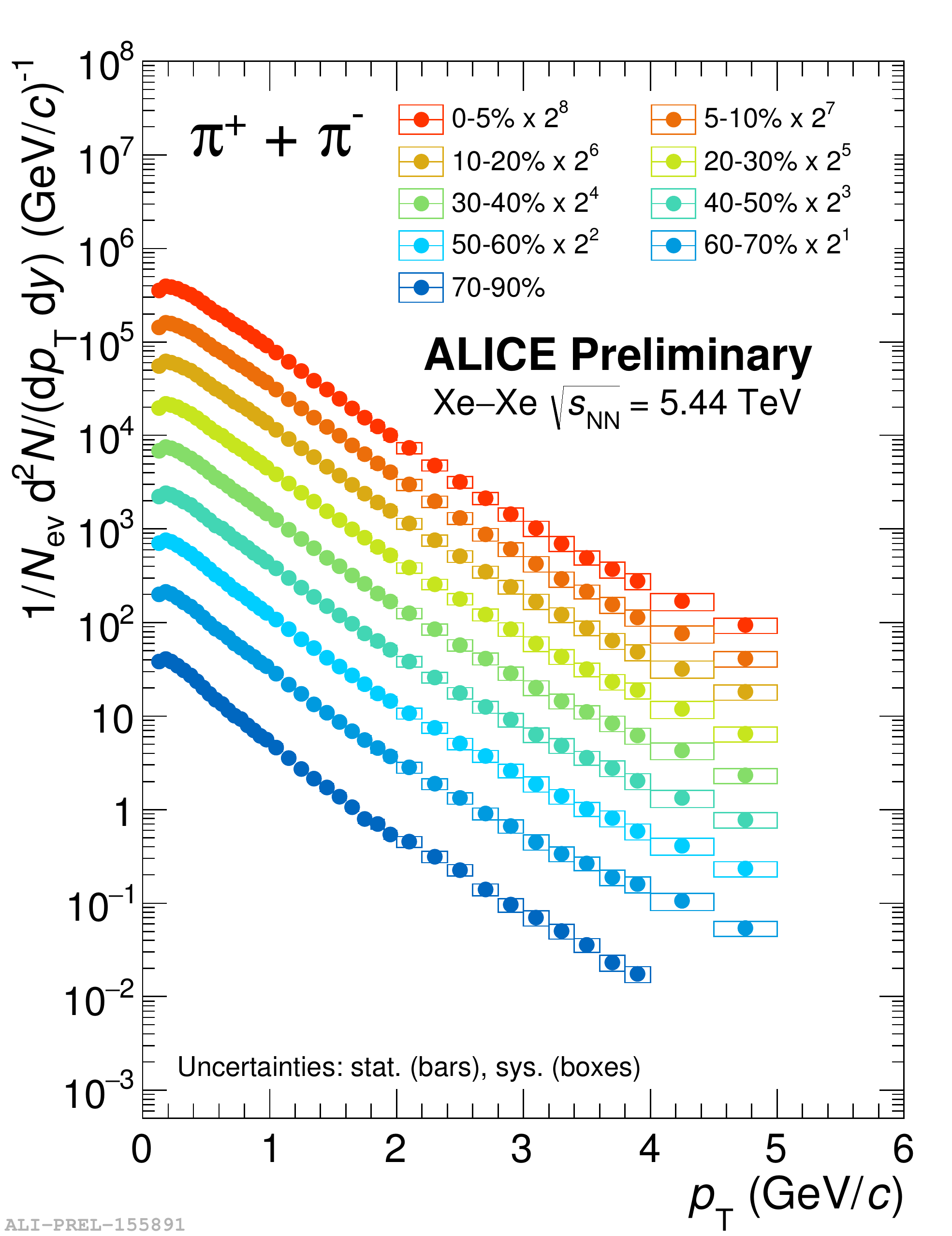}
      \label{fig:combixexe_pion}
  \end{subfigure}
  \begin{subfigure}{0.3\textwidth}
      \centering
      \caption{Combined spectra for K.}
      \includegraphics[width=\linewidth]{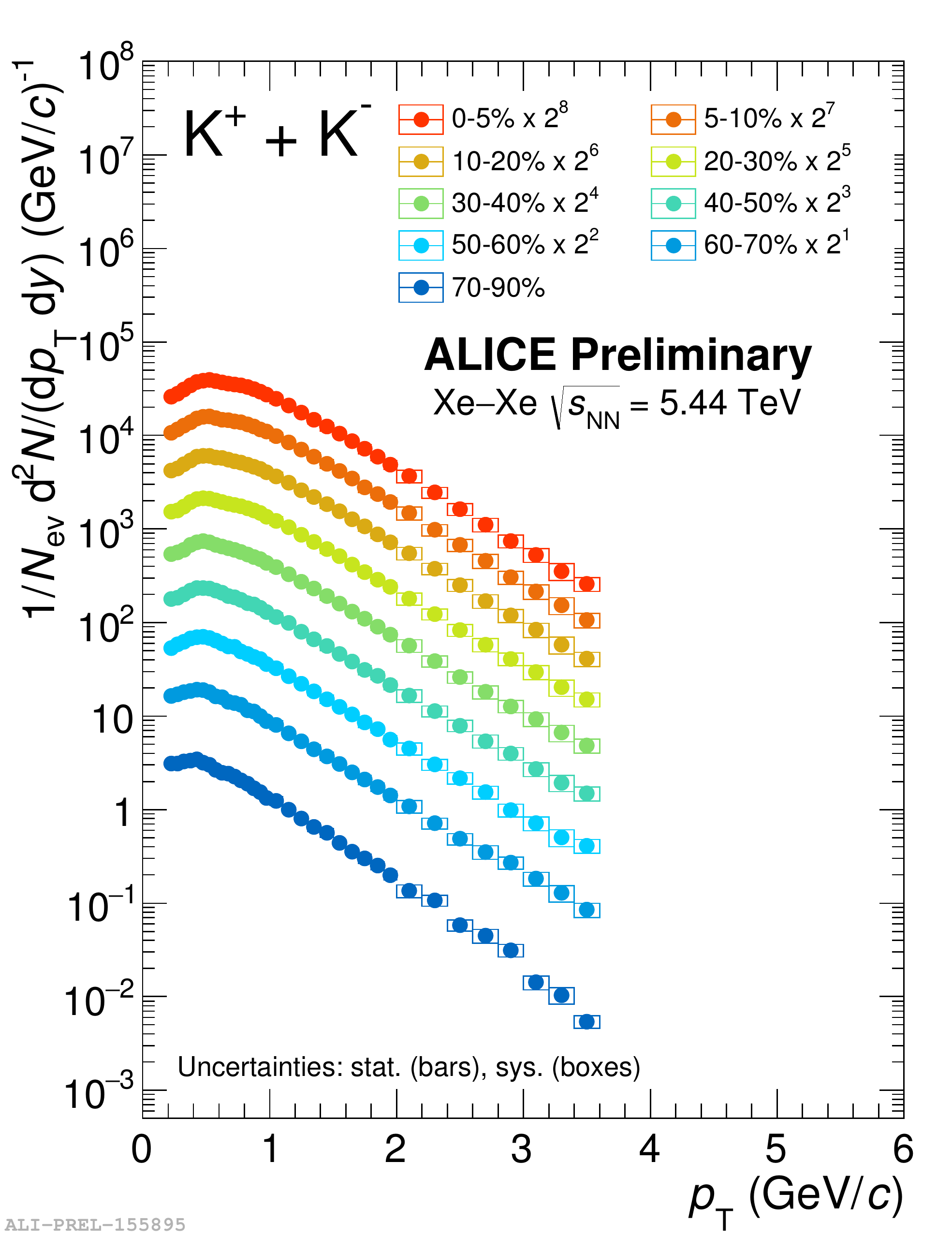}
      \label{fig:combixexe_kaon}
  \end{subfigure}
  \begin{subfigure}{0.3\textwidth}
      \centering
      \caption{Combined spectra for p.}
      \includegraphics[width=\linewidth]{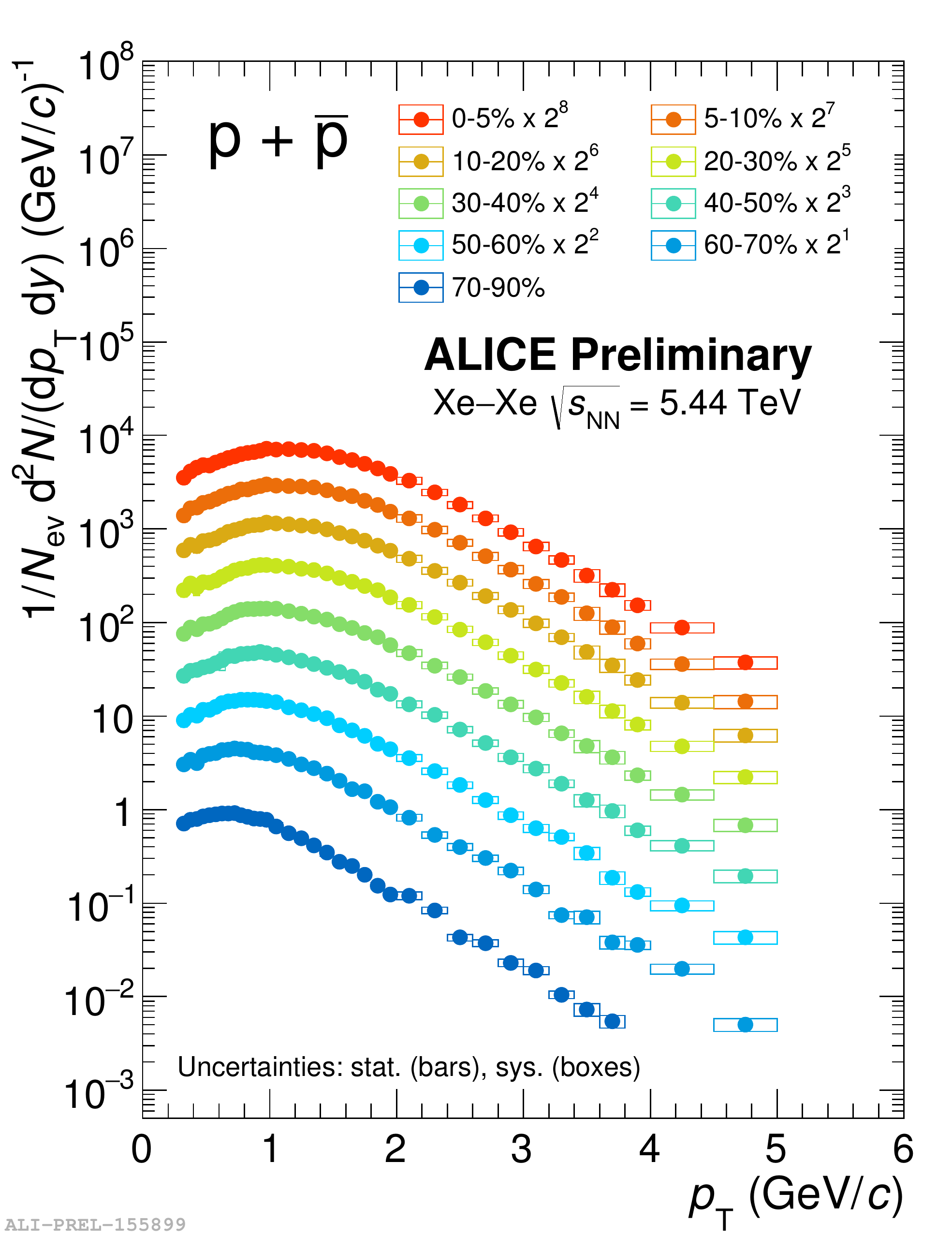}
      \label{fig:combixexe_proton}
  \end{subfigure}
    \caption{Combined spectra for $\pi$, K and p for all the centrality classes 
    considered in Xe--Xe collisions at $\sqrt{s_{\rm NN}} = 5.44 \text{ TeV}$. 
   }
\end{figure}

The combined spectra are shown on 
Fig.~\ref{fig:combixexe_pion},~\ref{fig:combixexe_kaon}, and 
\ref{fig:combixexe_proton}.
 It is evident that they show all the features expected from radial flow: the 
spectra get harder with increasing mass of the particle of interest, e.g. going 
from pions to protons, and the spectra get harder with increasing centrality 
(most evident for the proton spectra). The $p_{\rm T}$ spectra here shown, may 
be then fitted with a Blast-Wave parameterization, in order to extract the  
kinetic freeze-out temperature and the radial flow velocity. Subsequently, the 
values obtained for Xe--Xe collisions, may be compared with those obtained for 
Pb--Pb 
collisions at $\sqrt{s_{\rm NN}} = 2.76\text{, }5.02 \text{ TeV}$.


\begin{thebibliography}{99}
  \bibitem{checks_10}{P.J.~Siemens and J.O.~Rasmussen: Phys.~Rev.~Lett. 42, 880 
  (1979) 157, 158.}
  \bibitem{checks_11}{E.~Schnedermann, J.~Sollfrank and U.W.~Heinz: 
  Phys.~Rev.~C 
  48, 2462 (1993) 157, 158.}
  \bibitem{detector}{B.~Abelev et al. (ALICE Collaboration), Int.~J.~Mod.~Phys. 
  A29 (2014) 1430044.}
    \bibitem{time_event}{J.~Adam et al. (ALICE
    Collaboration), Eur.~Phys.~J.~Plus 132 (2017) no.2, 99. }

\end{thebibliography}
\end{document}